\documentclass[numberedappendix,apj]{emulateapj-rtx4}

\def\heao{{\it HEAO 1\/}}

\def\ltsima{$\; \buildrel < \over \sim \;$}
\def\simlt{\lower.5ex\hbox{\ltsima}}
\def\gtsima{$\; \buildrel > \over \sim \;$}
\def\simgt{\lower.5ex\hbox{\gtsima}}
\def\kms{\ifmmode{~{\rm km~s^{-1}}}\else{~km s$^{-1}$}\fi}
\def\lsim{\lower0.3em\hbox{$\,\buildrel <\over\sim\,$}}
\def\gsim{\lower0.3em\hbox{$\,\buildrel >\over\sim\,$}}

\def\h2{H$_2$}

\def\heao1{{\it HEAO-1\/}}

\begin{document}
\title{The hard X-ray perspective on the soft X-ray excess}
\author{Ranjan V. Vasudevan\altaffilmark{1,*}, Richard F. Mushotzky\altaffilmark{1,2}, Christopher S. Reynolds\altaffilmark{1,2} \\ Andrew C. Fabian\altaffilmark{3}, Anne M. Lohfink\altaffilmark{1}, Abderahmen Zoghbi\altaffilmark{1,2}, Luigi C. Gallo\altaffilmark{4}, Dominic Walton\altaffilmark{5}}

\altaffiltext{1}{Department of Astronomy, University of Maryland, College Park, MD 20742-2421}
\altaffiltext{2}{Joint Space-Science Institute (JSI), College Park, MD 20742-2421}
\altaffiltext{3}{Institute of Astronomy, Madingley Road, Cambridge, CB3 0HA, UK}
\altaffiltext{4}{Department of Astronomy and Physics, Saint Mary's University, 923 Robie Street, Halifax, Nova Scotia, B3H 3C3, Canada}
\altaffiltext{5}{Cahill Centre for Astronomy and Astrophysics, California Institute of Technology, Pasadena, CA 91125}

\altaffiltext{*}{ranjan@astro.umd.edu}

\begin{abstract}

The X-ray spectra of many active galactic nuclei (AGN) exhibit a `soft excess' below 1~keV, whose physical origin remains unclear.  Diverse models have been suggested to account for it, including ionised reflection of X-rays from the inner part of the accretion disc, ionised winds/absorbers, and Comptonisation.  The ionised reflection model suggests a natural link between the prominence of the soft excess and the Compton reflection hump strength above 10~keV, but it has not been clear what hard X-ray signatures, if any, are expected from the other soft X-ray candidate models.  Additionally, it has not been possible up until recently to obtain high-quality simultaneous measurements of both soft and hard X-ray emission necessary to distinguish these models, but upcoming joint \emph{XMM-NuSTAR} programmes provide precisely this opportunity.  In this paper, we present an extensive analysis of simulations of \emph{XMM+NuSTAR} observations, using two candidate soft excess models as inputs, to determine whether such campaigns can disambiguate between them by using hard and soft X-ray observations in tandem.  The simulated spectra are fit with the simplest ``observer's model'' of a black body and neutral reflection to characterise the strength of the soft and hard excesses.  A plot of the strength of the hard excess against the soft excess strength provides a diagnostic plot which allows the soft excess production mechanism to be determined in individual sources and samples using current state-of-the-art and next generation hard X-ray enabled observatories. This approach can be straightforwardly extended to other candidate models for the soft excess.

\end{abstract}

\section{Introduction}
\label{sec:intro}

Matter accreting onto supermassive black holes produces emission from radio--to--X-ray energies, with a variety of different physical processes being relevant in each band.  The direct luminous thermal power output from the central accretion disc appears in the optical--UV regime; a corona of electrons above the accretion disc Compton up-scatters this radiation to produce a power-law X-ray tail extending to hundreds of keV or greater depending on the coronal temperature.  This ensemble is often shrouded by dust and gas which introduces absorption features including a low-energy fall-off from photoelectric absorption, and troughs/edges from ionised absorption. Fall-offs are observed at higher energies due to Compton-scattering, and X-ray reflection from the accretion disc can reprocess coronal emission to produce a distinctive set of continuum features, including a reflection hump peaking at 30~keV and a broad Iron line at or near 6.4~keV.

Many AGN exhibit an excess at soft X-ray energies ($\lesssim 2$~keV) over the coronal power-law.  This was first noticed by \cite{1980ApJ...241L..13H} and \cite{1981ApJ...251..501P}, and more definitively confirmed with broad-band observations by \cite{1985ApJ...297..633S} and \cite{1985MNRAS.217..105A}.  The feature can be modelled as a black body with an almost universal effective temperature of $\sim$0.1--0.2~keV.  The physics of this feature remains uncertain, despite it contributing a potentially significant fraction of the total luminous power output of AGN.  The small range of effective temperatures observed disfavours a scenario where it represents the hard tail of the accretion disc spectrum peaking in the far-UV (e.g., \citealt{2009MNRAS.394..443M,2010MNRAS.406.2591P}), as sources with different masses and accretion rates would be expected to exhibit soft excesses with different temperatures.  Instead, atomic process are often invoked to explain the soft excess, whereby a series of lines are smeared together to produce the feature.  Two mechanisms can be invoked here: 1) ionised reflection with light bending, in which the series of soft X-ray lines are relativistically blurred due to being produced very close to the centre of the accretion flow of a rapidly-spinning black hole \citep{2005MNRAS.358..211R,2006MNRAS.365.1067C,2013MNRAS.428.2901W}, or 2) ionised absorption, where the soft excess is not really an `excess' at all, but the remnants of the true power law at low energies, with the high energy power-law being absorbed by smeared, ionised absorption with very high cloud velocities \citep{2004MNRAS.349L...7G}.  

Both models produce statistically acceptable fits to \emph{XMM} data for the same set of PG quasars (\citealt{2007MNRAS.381.1426M}, \citealt{2006MNRAS.365.1067C}), but both required strong relativistic smearing of either the emission or absorption features. \cite{2008MNRAS.386L...1S} found that this was problematic in the absorption model case due to the extreme terminal velocities required for the outflows; however, the model can still explain the data if the absorption is partially covering or clumpy.  In order to naturally produce a smooth soft excess with ionised reflection, one requires strong relativistic smearing from a rapidly spinning black hole, smoothing the soft features, and light bending from a corona with variable height above the accretion disc also allows the strength of the reflected emission (and thus of the soft excess) to vary relative to the powerlaw, and even to dominate at times. This is also supported by the observation of a time lag between the soft excess and direct continuum in the Narrow-Line Seyfert 1 galaxy 1H 0707-495 \citep{2009Natur.459..540F,2010MNRAS.401.2419Z} and in a sample of 32 AGN \citep{2013MNRAS.431.2441D}. Other processes can also account for the soft excess, such as Comptonisation of inner-disc photons (e.g., Done et al. 2012), and most recently, magnetic reconnection has also been suggested \citep{2013ApJ...773...23Z}.  It is not certain, using soft-X-ray observations alone, what process is responsible for producing the soft excess (\citealt{2006A&A...449..493C}, \citealt{2008A&A...482..499D}, \citealt{2013ApJ...777....2L}).   In this paper, we outline the different hard X-ray signatures that two physically-motivated models are expected to produce, and outline how using relatively simple fitting methods that do not require very high signal-to-noise ratio data, one can begin to distinguish between these models using the present and future generation of X-ray observatories.

Understanding the soft excess is important for determining the true bolometric output from accretion, and potentially provides a perspective on the extreme physical environments in AGN.  If light-bending and ionised reflection are important, it implies that the soft excess constitutes a real component of the ionising luminosity from the central engine and requires rapidly spinning black holes (e.g. \citealt{2013MNRAS.428.2901W}); whereas if ionised absorption is more relevant in some sources, it implies that the soft `excess' belies a much larger amount of ionising flux which we do not see due to high-energy absorption.  These two scenarios can drastically change the inferred bolometric luminosity from the central engine.  The implications for the bolometric luminosity from other models are less clear, highlighting the pressing need to understand how the soft excess is produced.

In the ionised reflection model for the soft excess, we expect to see an accompanying hard ($>$10~keV) excess, but simultaneous measurements of the soft and hard bands exist only for a few sources.  Hard X-ray observatories such as \emph{RXTE} and \emph{Swift/BAT} have revealed such hard excesses in many AGN, typically a smooth feature that can be produced by reflection from either the accretion disc (ionised) or more distant neutral material (e.g. the inner edge of the putative `torus', or in general, clouds of absorbing gas surrounding the AGN).  The overall strength of this `Compton hump' can be measured simply using the reflection parameter $R$ of the {\sc pexrav} model \cite{1995MNRAS.273..837M} available in the X-ray spectral fitting package {\sc xspec} \citep{1996ASPC..101...17A}. The value of $R$ in this model represents the total contribution from both distant and ionised reflection \citep{2007MNRAS.382..194N,2010MNRAS.408..601W,2013ApJ...772..114R}, although complex, Compton-thick absorbers may also be able to explain the hard excess \citep{2013ApJ...762...80T,2013ApJ...773L...5M,2009ApJ...698...99T}.  Disentangling the components from ionised and neutral reflection (e.g. \citealt{2007MNRAS.382..194N}) is important in trying to understand the soft excess, since it can only be produced by heavily blurred ionised reflection, not reflection from neutral material.  However, this requires very good signal-to-noise ratio data and is only possible for a few tens of sources currently.  To add confusion, ionised absorption can also produce a rising spectrum above $\sim$7~keV until about 20~keV similar to that seen from a Compton reflection hump.  It is only with the recent advent of \emph{NuSTAR} (launched 2012, \citealt{2013ApJ...770..103H}), that we have real prospects to break the degeneracy between these different models, as do future hard X-ray sensitive missions such as \emph{ASTRO-H} \citep{2012SPIE.8443E..1ZT} and \emph{ASTROSAT} \citep{2013IJMPD..2241009P}. 

In this paper, we ask whether the current state-of-the-art \emph{NuSTAR} mission, with good sensitivity up to $\sim$50~keV, in combination with the excellent 0.4--10~keV sensitivity of \emph{XMM}, can be used to distinguish between these two scenarios using only simple characterisations of the soft and hard excess strengths, using only short ($\sim$10~ks), inexpensive `snapshot' observations.

This paper is organised as follows: in section \ref{bathints} we describe preliminary evidence for a link between soft excess and hard excess strengths; in section \ref{sec:simulations} we describe the simulations of this relationship using ionised reflection and absorption models; in section \ref{sec:discussion} we discuss the results of those simulations and in section \ref{sec:summary} we summarise and discuss the implications of our findings.

\section{Hints of an $R-S_{\rm softex}$ relation from the BAT AGN catalogue}
\label{bathints} 

\begin{figure}
\centerline{
\includegraphics[width=9.0cm]{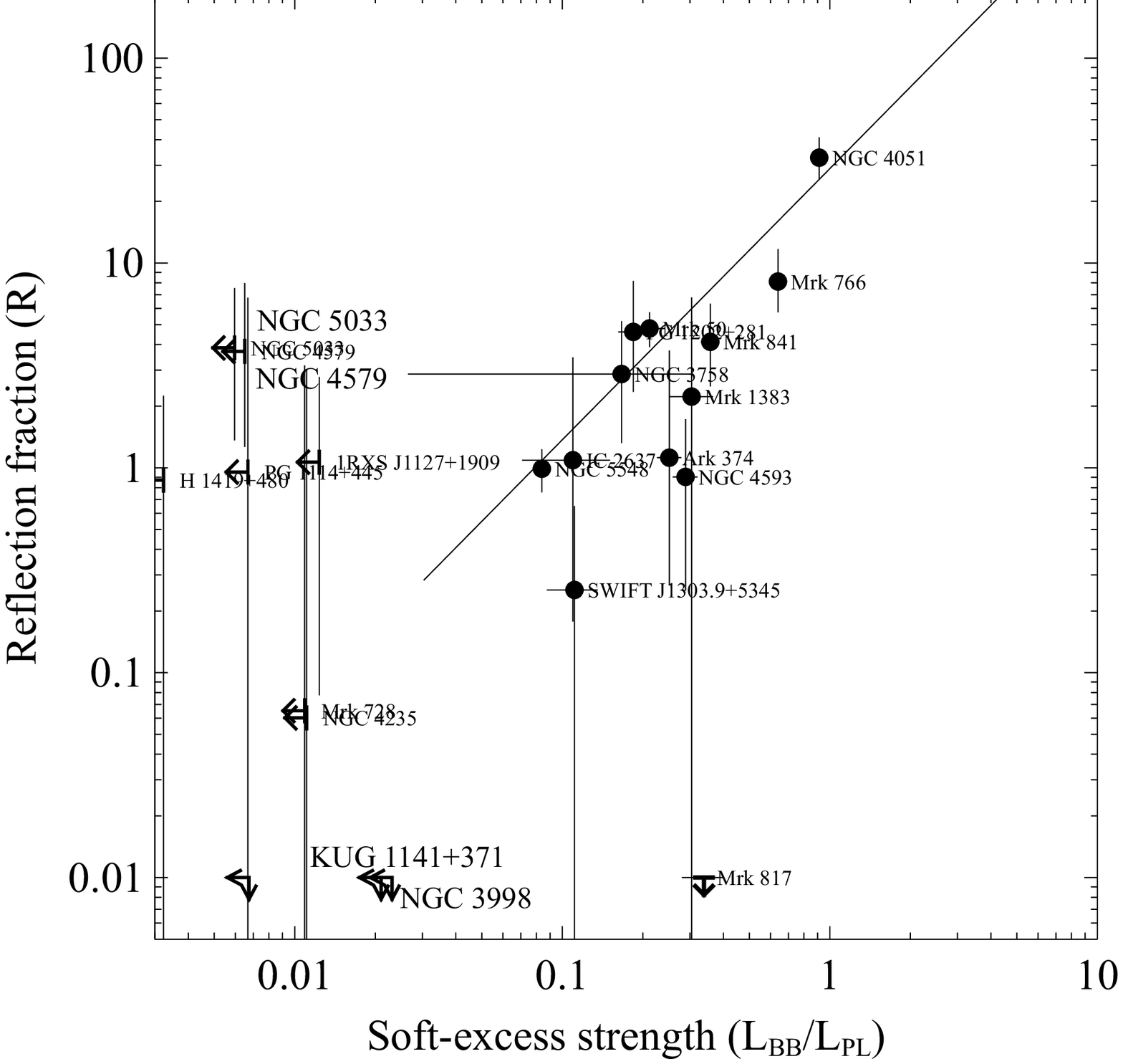}
}
\caption{Reflection vs. soft excess strength measured in the Northern Galactic Cap unabsorbed ($\rm log \thinspace N_{\rm H} \thinspace <22$) BAT AGN.  The solid line shows a simple best-fit to the points with well-detected soft excesses and hard X-ray reflection excesses.
\label{reflvssoftex_BAT58monAGN}}
\end{figure}

In \cite{2013ApJ...763..111V} (V13 hereafter), we present a comprehensive analysis of the 100 Northern Galactic Cap AGN ($b>50^{\circ}$) in the 58-month BAT catalogue.  The \emph{Swift/BAT} instrument provides the most unbiased census of AGN due to its sensitivity in the 14--195~keV band, less prone to the effects of absorption and host-galaxy diluation than any other band, and the Northern Galactic Cap provides a complete, manageable subsample of the catalogue from which to draw statistically robust conclusions on the AGN population.  In V13, BAT and \emph{XMM} data were used to constrain the reflection and soft excess strengths in 39 of these sources, including upper limits where such features were not detected.  The soft excess strength is presented there as the ratio of the luminosity of the feature (using a black body to model it) from 0.4--3~keV, to the luminosity in a relatively `clean', feature-free portion of the primary power-law from 1.5--6~keV.  We present below a plot (Fig.~\ref{reflvssoftex_BAT58monAGN}) of the reflection strength $R$ against the soft excess strength $S_{\rm softex}=L_{\rm BB}/L_{\rm PL}$ for those 23 low-absorption sources from V13 ($\rm log \thinspace N_{\rm H} < 22$), in which a soft excess could be detected if present.  There are strong hints of a correlation between $R$ and soft excess strength, but there are also many upper limiting reflection strengths or soft excess strengths which do not suggest any correlation.  This may suggest that different physical mechanism may account for these two families of sources.  However, the results using \emph{XMM} and BAT need to be treated with caution: firstly the BAT data are averaged over many months whereas the \emph{XMM} data represent snapshots over a few tens to hundreds of kiloseconds; they are therefore not simultaneous in any way, and as is known from NGC 4051 \citep{2006MNRAS.368..903P} and Mrk 590 \citep{2012ApJ...759...63R}, both the measured soft excess strength and reflection can vary significantly between observations separated by a few months-to-years.  The cross-normalisation between the hard and soft-band is therefore uncertain.  For 40\% of the sources in V13, the soft-band data (0.4--10~keV) were taken during the time frame of the BAT survey, so the BAT light curves have been used to re-normalise the BAT spectra to the level appropriate for the timeframe at which the soft-band data were taken.  This changes the measured reflection significantly in some sources, due to the changed relative flux at hard and soft energies. However, this only takes into account variation in the absolute normalisation of the spectra, not the spectral shape, which may also vary, and is only possible for the 40\% of sources for which the soft-band data were taken within the time frame of the BAT survey.  Therefore, using \emph{XMM} and BAT data together still leaves considerable uncertainties in the determination of the reflection strength $R$.

Since it is not robust to make this comparison using \emph{XMM} and BAT data alone, simultaneous broad-band data are preferred. This is available from joint \emph{XMM}-\emph{NuSTAR} campaigns, where the overlap between the \emph{NuSTAR} and \emph{XMM} bands allows the cross-normalisation to be constrained very well.  Such capabilities will also be offered by \emph{ASTRO-H} and \emph{ASTROSAT}.  However, before embarking on observing programmes with these missions, it is necessary to understand what the current favoured models for the soft excess would predict for the relation between the soft excess strength and the measured hard excess strength.  Regardless of the underlying physics responsible behind any given source spectrum, the observer's typical first-pass model will likely be a simple model such as a power-law with a black body, and a single reflection component at high energies (modelling both distant and inner-disc reflection).  Such a model has fewer degrees of freedom than complex ionised reflection or absorption models, and is more readily constrained by the type of data that are typically available for most sources.  In this paper, we simulate what the more complex models would look like if observed jointly by \emph{XMM} and \emph{NuSTAR}, and then fit the simplest ``Observer's Model'' combination of a {\sc pexrav} and black body model to measure the reflection hump and soft excess, respectively.  We then investigate whether the results resemble the tentative hints of the $R$-soft excess strength relation seen in the BAT sample (Fig.~\ref{reflvssoftex_BAT58monAGN}), and discuss whether \emph{XMM}-\emph{NuSTAR} campaigns can distinguish between different models in $R-S_{\rm softex}$ space.

\section{Simulations}
\label{sec:simulations}

\subsection{Ionised reflection}
\label{subsec:ionrefl}

We perform all of our simulations using the {\sc xspec} package \citep{1996ASPC..101...17A}.  The {\sc `fakeit'} command in {\sc xspec} provides the ability to generate a simulated spectrum for a given input model and instrumental response.  We first create a grid of simulated spectra for the ionised reflection case, stepping through all available parameters over physically realistic ranges based on previous studies using this model.  The model combination used for generating simulated spectra is {\sc pexrav + kdblur(reflionx)}.  The key component is {\sc reflionx} \citep{2005MNRAS.358..211R}, which provides all the essential features of the reflected continuum including the Compton reflection hump, broad Iron line and soft lines which are thought to be blurred to give rise to a soft excess.  This model takes the following parameters: the Iron abundance relative to solar Iron abundance, $A_{\rm Fe}$ (we step this between 0.11 and 3.0, the range allowed by the model); the photon index of the illuminating coronal spectrum being reflected from the disc, $\Gamma_{\rm refl}$ (for an input photon counts spectrum $N(E) \propto E^{-\Gamma_{\rm refl}}$ - we step this between 1.5 and 3.0 (e.g. V13, \citealt{2011A&A...530A..42C}); the ionisation parameter $\xi$ (the ratio of the illuminating flux to the hydrogen number density, stepped between 1 and 1000, e.g. \citealt{2008ApJ...675.1048R}) and the normalisation $N_{\rm reflionx}$.  We blur this component with the {\sc kdblur} model, which requires a radial emissivity index of the disc (we set this at $5.0$, to allow for modest light-bending), the inner and outer radii of the disc ($R_{\rm in}$ and $R_{\rm out}$, here frozen at $3.0 R_{\rm g}$ and $100.0 R_{\rm g}$ respectively, again to allow for relativistic effects from a spinning black hole allowing an innermost stable orbit within 6 $R_{\rm g}$), and the inclination (which we leave at a default of $30^{\rm \circ}$ for all realisations, assuming relatively face-on, Seyfert-1--like geometry).   

The final component of the model combination to be considered is {\sc pexrav}, which in our representation here represents the direct or illuminating power-law coronal continuum, along with simple reflection from distant, neutral material (e.g. the inner surface of the surrounding dusty gas clouds or `torus').  The {\sc pexrav} model requires the photon index $\Gamma_{\rm direct}$, the cut-off energy (high energy fall-off) of the spectrum $E_{\rm cut}$, the reflection fraction $R_{\rm distant}$, the abundances of Iron and other metals $A_{\rm Fe,distant}$ and $A_{\rm other,distant}$, the cosine of the inclination angle and the normalisation ($N_{\rm pexrav}$). We link the photon index of the direct component to the reflected component ($\Gamma_{\rm direct}=\Gamma_{\rm refl}$) and freeze the cut-off energy at the maximal value $10^{6}$~keV, since current BAT-based studies suggest that the average cut-off energy for AGN consistently lies above a few hundred keV in the majority of AGN (37 out of 49 AGN fit with {\sc pexrav} in V13 AGN have $E_{\rm cut}$ values consistent with being outside the BAT band, or are poorly constrained); at any rate we make the assumption that it lies outside the \emph{NuSTAR} band, to consider the simplest case first in our simulations.  We freeze the abundances at their default values (these represent the abundances of the distant reflector) and keep the inclination angle at its default value.  \cite{2007MNRAS.382..194N} de-convolve distant and inner reflection in a sample of Seyfert AGN, and on average find that the distant component of the reflector has a strength $R_{\rm distant} = 0.455$ with a standard deviation $\sigma_{\rm R,dist} = 0.295$.  For each realisation of the input spectrum, we randomly generate a Gaussian-distributed distant reflection value using these parameters, to introduce a realistic amount of spread due to the presence of some distant reflection.

Lastly, the crucial parameter in this model set-up is the ratio of the normalisations of the direct and reflected components.  Before addressing this, we first need to understand a complication presented by the published version of the {\sc reflionx} model: changing the ionisation parameter actually changes the flux of the source in addition to changing the spectrum, so the normalisation is not a simple flux-multiplier.  We modify the table model such that the normalisation is divided out from the ionisation parameter, and variations in $\xi$ only produce spectral shape changes (not overall flux changes).  Having made this change, we can use the normalisation of the {\sc reflionx} model as a simple flux multiplier.  We allow the ratio $N_{\rm pexrav}/N_{\rm reflionx}$ to go between 1.0 (representing the reflection-dominated case) and 1000.0 (where the power-law dominates and ionised reflection signatures should be barely discernible in the spectrum).  We finally set the overall 1--200~keV flux of the input model to be $4 \times 10^{-11} \rm erg \thinspace s^{-1} \thinspace cm^{-2}$, which is the measured flux for the well-known Seyfert NGC 4051 using \emph{XMM} and BAT observations, and the flux of a typical \emph{NuSTAR} target.

Using the above set-up, we now simulate spectra in both the \emph{XMM} and \emph{NuSTAR} bands.  We employ more `pessimistic' assumptions for the simulated spectra and just simulate spectra for the PN instrument from \emph{XMM} and the FPMA instrument from \emph{NuSTAR}; if more detectors are used in data fits, then the accuracy of results obtained should only increase.  For the simulated \emph{XMM} spectrum, we use the response and ancillary files from an observation of NGC 4051 reduced in V13, and for the \emph{NuSTAR} simulated spectrum, we use the latest available canned response files for the FPMA instrument.  {Since both the \emph{NuSTAR} and \emph{XMM} responses oversample the spectrum, we re-bin the responses for quicker fits of the simulated spectra.  This was done using the {\sc ftools} tasks {\sc rbnrmf} and {\sc marfrmf} for re-binning the response by user-specified amounts in each group of channels, and finally combining the effective area file and response file.   We again assume a relatively conservative exposure time of 10 ks in both \emph{XMM} and \emph{NuSTAR}, typical of the snapshots of BAT AGN  currently being taken by \emph{NuSTAR}; longer observations will provide still better constraints on the soft and hard excess strengths.  However, the key point of this study is to see whether physical models for the soft excess can be distinguished using broad-band X-ray data of \emph{typical} quality rather than using the longest observations available, and using \emph{simple} models rather than the more complex, physically motivated models.  We present an example of the model spectrum and resultant simulated spectrum in Fig.~\ref{fig:example_modelandfake_ionisedrefl}.

\begin{figure}
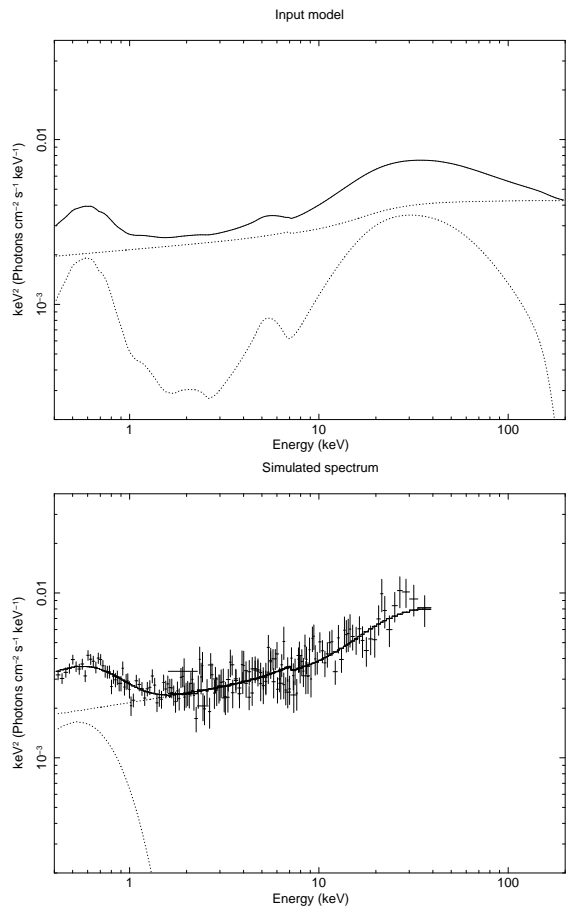


\includegraphics[width=6.0cm,angle=270]{examplemodelspec_ionisedrefl.eps}
\includegraphics[width=6.0cm,angle=270]{examplefakedspec_ionisedrefl.eps}

\caption{Model spectrum (top panel) and resultant simulated \emph{NuSTAR} and \emph{XMM} spectra (lower panel), using ionised reflection as the input model.  The simulated spectrum is fit with the simplest ``observer's model'' combination {\sc pexrav + gauss + bbody}.  Short exposure times result in few counts above 50~keV.
\label{fig:example_modelandfake_ionisedrefl}}
\end{figure}

Having simulated the spectra, we group them both using the {\sc grppha} tool to have a minimum of 20 counts per bin and re-load the binned spectra into {\sc xspec}. We ignore any data outside of the range 0.4--10~keV in the simulated \emph{XMM} data and any data above 80~keV and below 3~keV in the simulated \emph{NuSTAR} data, as well as ignoring any `bad' channels.  We then fit the spectra with what is henceforth referred to as the \emph{``Observer's model''}: the simplest combination of a black body, a gaussian and a {\sc pexrav} reflection model ({\sc bbody + gauss + pexrav}) required to account for the components seen in the spectrum.  We use initial conditions that mimic a typical soft-excess, power-law, iron-line and hard-excess spectral shape.  We set the normalisatoins of the black body, gaussian line and pexrav components to be $10^{-4}$,$10^{-4}$ and $10^{-2}$ respectively, and constrain the black body temperature to lie between 0.01 and 2~keV, the line energy to lie between 6.3 and 7~keV, and the linewidth to lie between 0 and 0.5~keV, as is commonly done when fitting real observations.  This provides a sensible starting point for the fit to ensure that the simulated spectra have a good probability of being fit successfully.

While the soft excess is probably not intrinsically black body emission and the hump above 10~keV may not intrinsically be due to reflection, these model components serve to parameterise the spectral shape in a way most commonly done by observers.  The purpose of the first component is to measure the strength of the soft excess when modelled as a black body, and to calculate its strength using the simple parameterisation $L_{\rm BB}/L_{\rm PL}$ introduced in V13.  The purpose of the gaussian component is to model the Iron line feature that is naturally introduced around 6.4~keV by the {\sc reflionx} model, and the purpose of the {\sc pexrav} component is to measure the \emph{overall} strength of the resultant Compton hump, \emph{which will in general be the sum of both distant and inner reflection}, or an artefact of complex absorption.  For data of typical quality, the abundances and inclination are not usually uniquely determinable so we do not thaw them here for fitting and freeze them at their defaults.  We also freeze the cut-off energy at the maximal $10^{6}$~keV in our Observer's model fit, since experience shows that this is rarely constrained to be below $\sim$200~keV in real AGN fits using BAT and \emph{XMM} data (V13).  We fit each simulated spectrum with this Observer's model, and record all the best-fit parameters (especially the measured soft excess and reflection strengths) for each set of input (simulated) parameters ($A_{\rm Fe}$, $\Gamma_{\rm direct}$, $\xi$ and the ratio of direct-to-reflected components $N_{\rm pexrav}/N_{\rm reflionx}$).

We step each parameter within the ranges indicated above, using 7 steps for four primary variables ($\xi$,$\Gamma$,$A_{\rm Fe}$ and $N_{\rm pexrav}/N_{\rm reflionx}$), amounting to 2401 total simulated spectra. After the fits to the simulated data are complete, we then plot the `measured' soft excess strength against the reflection to investigate the presence of any relation (Fig.~\ref{fig:refl_vs_softex_ionisedreflsimulations}).  We filter out poor fits to the simulated spectra using a dual measurement of the reduced $\chi^{2}$; requiring $\chi^{2} / \rm d.o.f < 6.0$ in the soft band ($<3.0$~keV) and $\chi^{2} / \rm d.o.f < 2.5$ in the hard band ($>10$~keV), to ensure that our Observer's model fits \emph{both} the soft excess and the hard excess simultaneously; a single band-wide reduced $\chi^{2}$ criterion was found to be insufficient to ensure this.  These reduced-$\chi^{2}$ thresholds were chosen by visual inspection of the spectral fits for different cut-off $\chi^{2}$ values, fine-tuning the thresholds until only those spectra with visually good fits to the soft and hard excesses remained.  We adopt different reduced-$\chi^{2}$ thresholds in the two bands as there are many more bins in the soft band than the hard band: it is therefore expected to be more difficult to get a good fit in the soft band using a simple two-parameter model. A total of 1962 of the 2401 fits (82\%) were deemed `good' fits according to these criteria. The `bad' fits represent parameter combinations for which the simple ``Observer's model'' could not adequately represent the soft and hard excesses. We omit error bars and do not calculate errors on individual simulated spectrum fits, but note that the error on $S_{\rm softex}$ for typical \emph{XMM}-quality data are very small, as shown in Fig.~\ref{reflvssoftex_BAT58monAGN}. We know \emph{NuSTAR} will constrain $R$ much more robustly than \emph{XMM}+\emph{BAT} fits; therefore errors on $R$ will assuredly be smaller than in Fig.~\ref{reflvssoftex_BAT58monAGN} and errors on $S_{\rm softex}$ will be comparable.

\begin{figure}
\centerline{
\includegraphics[width=9.0cm]{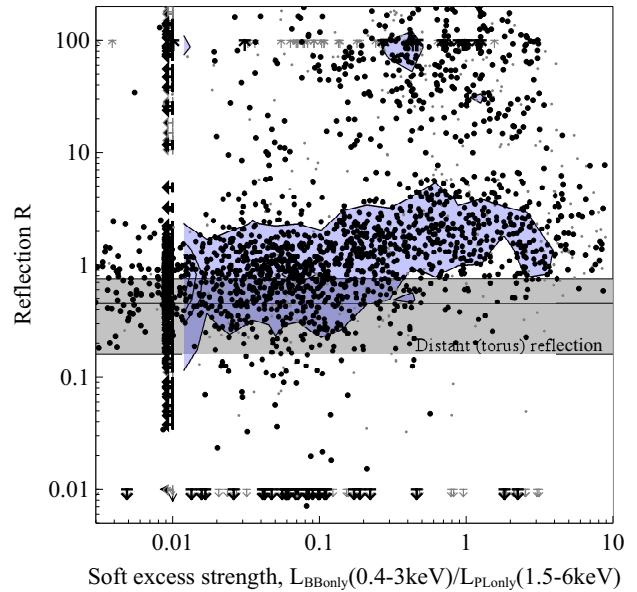}
}
\caption{Reflection vs. soft excess strength as measured from ionised reflection simulated spectra.  The smaller grey points show all results, and the black points show the results for which the Observer's model was a `good' fit to the simulated data, using the dual-$\chi^{2}$ criteria given in the text.  The contours show the clustering of the points, with the minimal contours representing 1/13th of the peak value at the centre of the contours.  The gray shaded area shows the expected 1-$\sigma$ range of reflection strengths from distant reflection from cold material, as found by \protect\cite{2007MNRAS.382..194N}.
\label{fig:refl_vs_softex_ionisedreflsimulations}}
\end{figure}

We see a large degree of spread in Fig.~\ref{fig:refl_vs_softex_ionisedreflsimulations}, but the contours of the highest density of points show a modest but clear trend of increasing $R$ with $S_{\rm softex}$.  Part of the region of high $R$ ($R \gtrsim 30$) shows a high density of points.  Performing a Kendall's-$\tau$ correlation analysis on the good fits only yields a correlation coefficient of 0.33 with a null-hypothesis probability $<1 \times 10^{-10}$.

\subsection{Ionised absorption}
\label{subsec:ionabs}

To illustrate the power of the $R-S_{\rm softex}$ diagnostic plot, we also perform this exercise for the ionised wind model, {\sc swind1}.  Although \cite{2008MNRAS.386L...1S} find this model to require unphysically high terminal velocities of the outflow, a partially-covering ionised absorber would resolve this problem, and therefore the {\sc swind1} model can still be taken as representative of an important class of multiple-absorber, partially-covering, ionised absorber models that can account for the soft excess. We therefore simulate spectra using this model, using the range of parameters identified in the \cite{2007MNRAS.381.1426M} study on PG quasars, to see whether such a model can produce soft `excesses' and hard `excesses' that can be modelled as a `black body + reflection' model combination.

We use the model combination {\sc swind1(pexrav)} to simulate spectra, where the {\sc pexrav} component represents the primary X-ray continuum along with some distant reflection.  In \cite{2007MNRAS.381.1426M}, a more complex model for the primary X-ray power-law and distant reflection is used, but the salient features of such a model are reproduced here by {\sc pexrav} for our purposes.  We follow exactly the same rationale for randomly seeding the distant reflection with values appropriate for the distribution found in \cite{2007MNRAS.382..194N} and follow identical rationale to that given above in \S\ref{subsec:ionrefl} for determining the primary continuum parameters; the primary continuum is common to both the ionised reflection and absorption cases.  For the ionised absorption component, we step the column density of the wind between 3 and 50 $\times 10^{22} \rm \thinspace cm^{-2}$; the photon index of the primary continuum ({\sc pexrav}) $\Gamma$ between 1.5 and 3; the logarithm of the ionisation parameter $\xi$ is varied between $2.1 < \rm log(\xi) < 4.0$; and the Gaussian velocity dispersion of the wind is varied between 0.1 and 0.5, all based on the parameters found from fitting data on real AGN in \cite{2007MNRAS.381.1426M}.  We simulate spectra in a grid using 7 steps between the maxima and minima for each parameter (see Fig.~\ref{fig:example_modelandfake_ionisedabs} for example simulated spectra), and plot the resulting soft excess strengths and reflection strengths in Fig.~\ref{fig:refl_vs_softex_ionisedabssimulations}. For this input model, 2073 out of 2401 potential parameter combinations yield successful ``Observer's Model'' fits (86\%); in the remainder of cases, a `black body plus reflection' model combination entirely failed to fit the simulated spectrum (i.e. {\sc xspec} could not compute a fit at all to produce a fit statistic).  Using the same dual reduced-$\chi^{2}$ criteria as for ionised reflection, we find that 1595 of the successful fits were `good' fits (i.e., 77\% of successful fits, or 66\% of the total 2401 simulated spectra).

We see a large degree of spread in Fig.~\ref{fig:refl_vs_softex_ionisedabssimulations}, without any clear trend linking $R$ with $S_{\rm softex}$.  Notably, a large number of simulated spectra show prominent soft excesses with negligible $R$.  Performing a Kendall's-$\tau$ correlation analysis on the good fits only yields a correlation coefficient of -0.003 with a null-hypothesis probability of 0.88, indicating that there is a high chance of the two properties being completely uncorrelated.

\begin{figure}
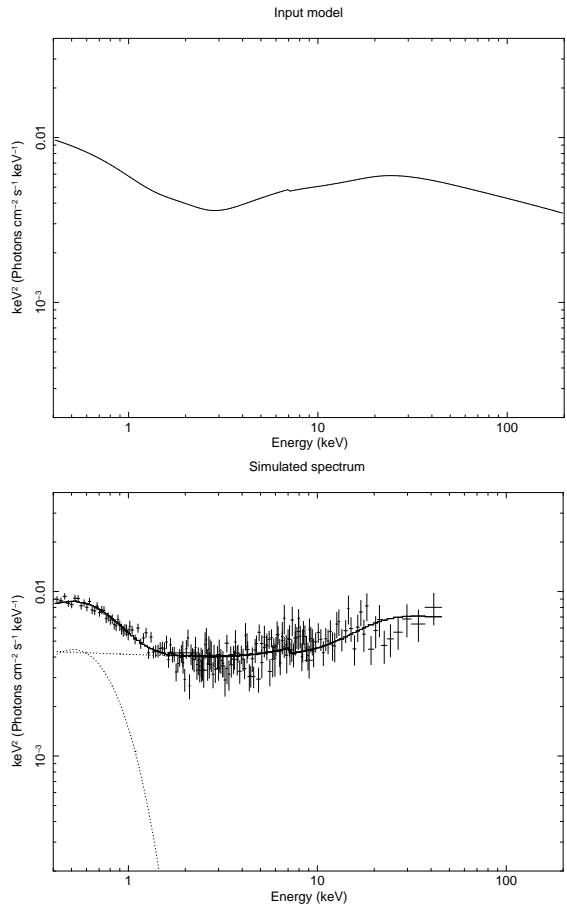


\includegraphics[width=6.0cm,angle=270]{examplemodelspec_ionisedabs.eps}
\includegraphics[width=6.0cm,angle=270]{examplefakedspec_ionisedabs.eps}

\caption{Model spectrum (top panel) and resultant simulated \emph{NuSTAR} and \emph{XMM} spectra (lower panel), using ionised absorption as the input model. The simulated spectrum is fit with the ``observer's model'' combination {\sc pexrav + bbody}. Short exposure times result in few counts above 50~keV.
\label{fig:example_modelandfake_ionisedabs}}
\end{figure}

\begin{figure}
\centerline{
\includegraphics[width=9.0cm]{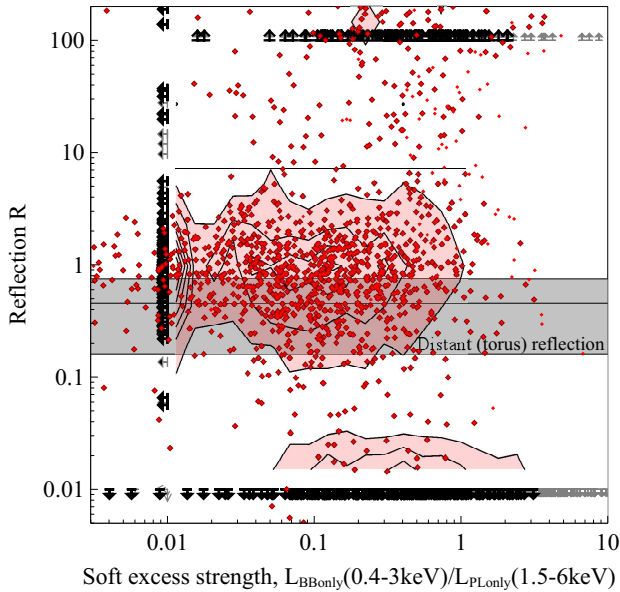}
}
\caption{Reflection vs. soft excess strength as measured from ionised absorption simulated spectra.  Key as for Fig.~\ref{fig:refl_vs_softex_ionisedreflsimulations}
\label{fig:refl_vs_softex_ionisedabssimulations}}
\end{figure}

\section{Discussion of Results}
\label{sec:discussion}

At the outset, we note that the ``Observer's Model'' produces a successful fit to all of the simulated spectra using ionised reflection as the input model, but only 86\% of the ionised absorption simulated spectra could be fit with this model.  This indicates that while ionised absorption can produce a wide range of spectral shapes that can mimic the appearance of a soft black body component with a hard X-ray reflection hump, a substantial minority do not.  Sources in those ranges of parameter space would exhibit spectral shapes clearly distinguishable from ionised reflection. We discuss this class of spectra further in \S\ref{subsec:failedabsfits}.

The two models compared here show some clearly distinct behaviour in $R-S_{\rm softex}$ space; the results for both models are plotted in Fig.~\ref{fig:refl_vs_softex_bothmodels_summary}.  Notably, in the ionised reflection scenario, stronger soft excesses can be produced and these are accompanied by stronger measured reflection fractions, particularly for $S_{\rm softex} \gtrsim 1$, $R \gtrsim 1$.  Ionised reflection as a mechanism for the soft excess could be distinguishable from distant reflection by a more pronounced hard excess (values of $R \gtrsim 2$) that one expects to obtain from such a physical process, coupled with a trend towards higher $R$ at higher $S_{\rm softex}$, which one would expect to observe in large samples of AGN.  This would imply that in sources where both 1) strong soft excesses and 2) strong reflection are measured, ionised reflection is the most likely candidate model.  For sources with $0.5 \lesssim R \lesssim 3$ and $S_{\rm softex} \lesssim 1$, the two models are not immediately distinguishable.

Ionised absorption, on the other hand, can produce hard excesses with strengths $R$ that are not easily distinguishable from distant (torus) reflection even if a strong soft excess is present.  This would imply that for sources with 1) a strong soft excess ($0.3<S_{\rm softex}<1$) but 2) a not particularly strong hard excess ($R \lesssim 1$), ionised absorption may be a candidate model to explain the soft excess, but it could also be due to an altogether different physical process, with an unrelated hard excess due to distant, neutral reflection.

\cite{2006A&A...449..493C} found that the strongest soft excesses can be produced by ionised absorption, not reflection, contrary to our findings here.  They investigated the different absorber conditions required to reproduce observed soft excesses in detail, alongside a blurred reflection model akin to the one we use here but with one key difference: the reflection spectrum strength is constrained to be sufficiently lower than that of the primary continuum such that prominent soft excesses cannot be seen.  In our model, the ratio of the direct {\sc pexrav} continuum to the {\sc reflionx} continuum is widely variable to account for light-bending effects (e.g., the coronal height varying above the accretion disc), and when $N_{\rm pexrav}/N_{\rm reflionx}\sim 1$, strong soft excesses can be observed, as seen in reflection-dominated epochs of sources such as NGC 4051 \citep{2006MNRAS.368..903P}.  Full consideration of the effects of light-bending at the inner parts of the accretion flow allows the reflection spectrum to dominate and produce such strong excesses.

Some very interesting behaviour is seen at the extrema of Fig.~\ref{fig:refl_vs_softex_bothmodels_summary}.  Both ionised reflection and absorption allow for non-detected soft excesses alongside moderate reflection ($0.2 < R < 3.0$), but ionised absorption allows for a large range of soft excess strengths alongside negligible or no measured hard excess.  Therefore, in sources where the soft excess is prominent but reflection is undetectable, it may be appropriate to consider partially-covering ionised absorption or other non-reflection based models (e.g., Comptonisation). Finally, the region of the plot showing strong soft excesses $S_{\rm softex}>1$ and relatively weak hard excesses $R<1$ is not populated by either of the two models considered here.  Further work needs to be done to explore if other models can occupy this part of parameter space.

We can apply these diagnostics now to the objects in Fig.~\ref{reflvssoftex_BAT58monAGN}, although with the caveat that those results were obtained using non-simultaneous hard- and soft-X-ray data, without the overlap between the soft and hard band required to constrain reflection well.  Nevertheless, one can suggest (subject to further investigation with better data) that those objects in which reflection and soft excess strength seem to be increasing together likely have a strong contribution from ionised reflection.  NGC 4051 has been studied in depth and it is known that ionised reflection can be fit to the detailed 0.4--10~keV spectrum (\citealt{2006MNRAS.368..903P}, \citealt{2013MNRAS.435.1511A}).  Reflection has been suggested for Mrk 766 also \citep{2011MNRAS.416L..94E}; but complex, multi-layered absorbers (perhaps a disk wind), and occultation by absorbing clouds have also been invoked \citep{2007A&A...475..121T,2011MNRAS.410.1027R}.  Both ionised reflection and absorption can explain the spectrum of Mrk 841 \citep{2011A&A...535A.113C}. The object Mrk 817, which has negligible reflection but a measurable soft excess, may well require a different model such as ionised absorption to account for its soft features.  Indeed, \cite{2011ApJ...728...28W} mention an epoch in this source where absorbing winds were detected in UV spectroscopy of this source from 1997 and 2009, albeit without the X-ray edges due to Oxygen expected at 0.73 and 0.87~keV from such outflows.  One potential explanation may be that the $>2$~keV X-ray continuum in this source is absorbed by a highly relativistic ionised absorber, which may tally with the absorption signatures in the UV.  Again, data that extends into the $>10$~keV band would provide more definitive answers, stressing the utility of the approach outlined in this paper.

\begin{figure}
\centerline{
\includegraphics[width=9.0cm]{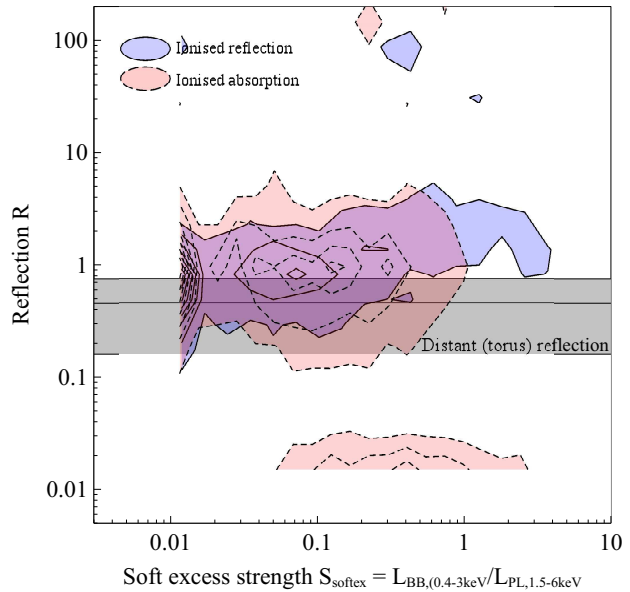}
}
\caption{Contours of reflection vs. soft excess strength from both ionised reflection and ionised absorption models, for comparison.
\label{fig:refl_vs_softex_bothmodels_summary}}
\end{figure}

Two Narrow-Line Seyfert 1 nuclei with strong soft excesses have recently been studied in detail using \emph{XMM} data: 1H 0707-495 and IRAS 13224-3809.  \cite{2012MNRAS.419..116F} and \cite{2012MNRAS.422.1914D} find that reflection can successfully fit the spectrum of 1H 0707-495; \cite{2013MNRAS.430.1408K} and \cite{2013MNRAS.429.2917F} find the same for IRAS 13224-3809.  We use the archival \emph{XMM} data to estimate $S_{\rm softex}$ for both of these sources and find that they are 3.3 (1H 0707-495) and 2.6 (IRAS 13224-3809); therefore according to the scheme found in this paper, their soft excesses are both sufficiently strong to favour a reflection-dominated scenario.  Broad-band observations with \emph{NuSTAR} should be able to confirm this and locate both objects on the $R-S_{\rm softex}$ plot.

\subsection{Extreme values of the hard excess strength, $R \gtrsim 100$}

For both models tested in this paper, there are a cluster of points at $R\gtrsim 100$, which have not so far been observed in the real AGN population.  For the ionised reflection case, further investigation reveals that all of the simulated spectra that produce $R>100$ have $N_{\rm pexrav}/N_{\rm reflionx} < 10$, and all of the very extreme $R$ values (i.e. $R > 200$) measured from `good' fits have $N_{\rm pexrav}/N_{\rm reflionx} = 1$, the lowest value of the ratio included in the simulations, corresponding to the most reflection-dominated case (see Fig.~\ref{fig:ionisedreflsplitbyNratioandxi}, top panel).  This suggests that for reflection-dominated spectra, the spectral shapes produced can genuinely give rise to very high, `anomalous' $R$ values (under the fitting assumptions adopted in this study), depending on the spread of other intrinsic parameters i.e. $A_{\rm Fe}$, $\xi$ and $\Gamma$.  We then focus only on the subset of objects with reflection-dominated spectra ($1 < N_{\rm pexrav}/N_{\rm reflionx} < 10$, Fig.~\ref{fig:ionisedreflsplitbyNratioandxi}, lower panel).   We split the results into different bins of $A_{\rm Fe}$, $\xi$ and $\Gamma$, selecting 3 bins for each (using logarithmic spacing for $\xi$).  We find that the ionisation parameter $\xi$ produces the greatest variation in measured $R$.  The lowest ionisation parameters, $1.0 < \xi < 10.0$, give rise to almost all of the $R>100$ points. In conclusion, the most reflection-dominated spectra ($1.0 < N_{\rm pexrav}/N_{\rm reflionx} < 10.0$) coupled with the most weakly ionised reflectors ($1.0 < \xi < 10.0$) can produce very strong hard excesses.

\begin{figure}
\includegraphics[width=8.0cm]{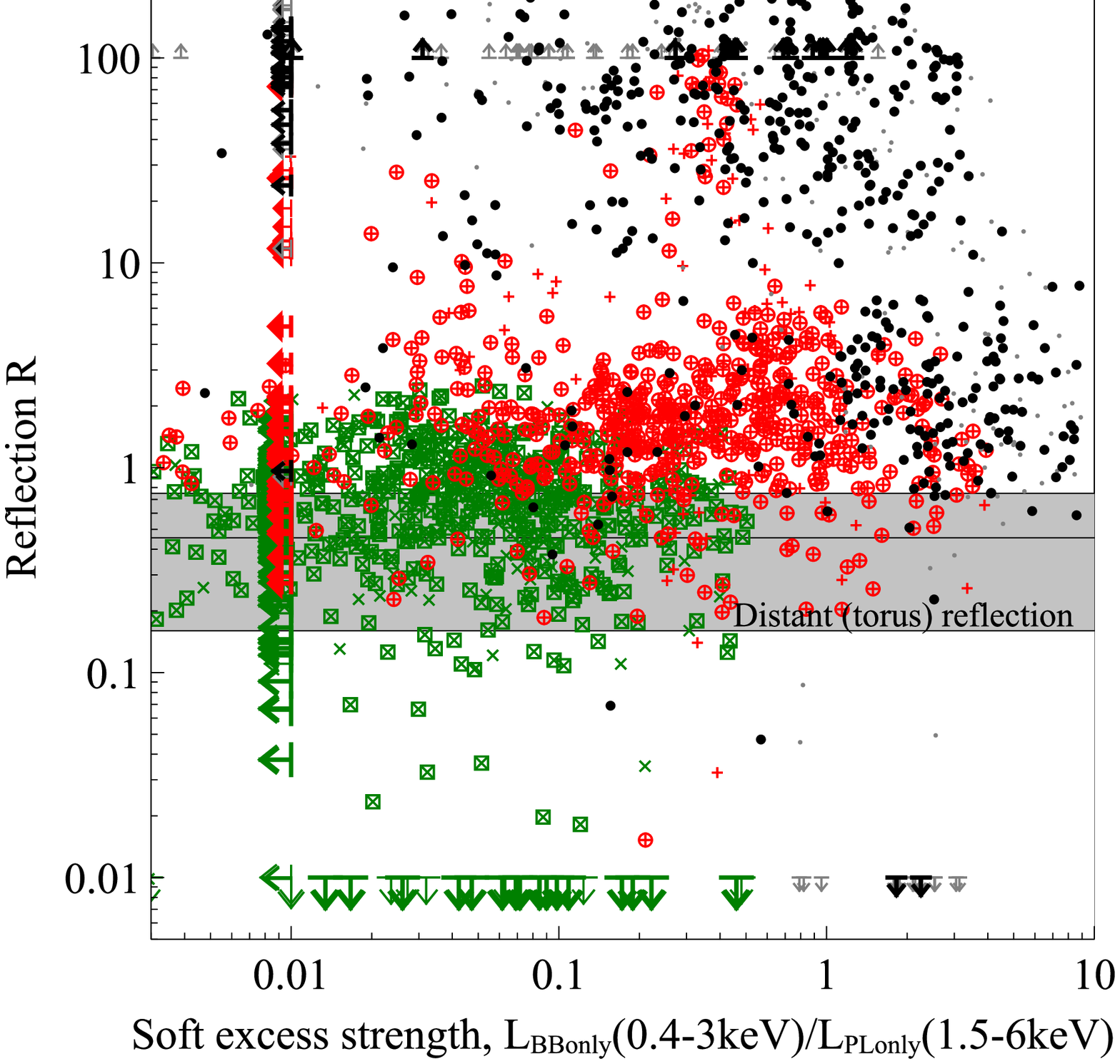}
\includegraphics[width=8.0cm]{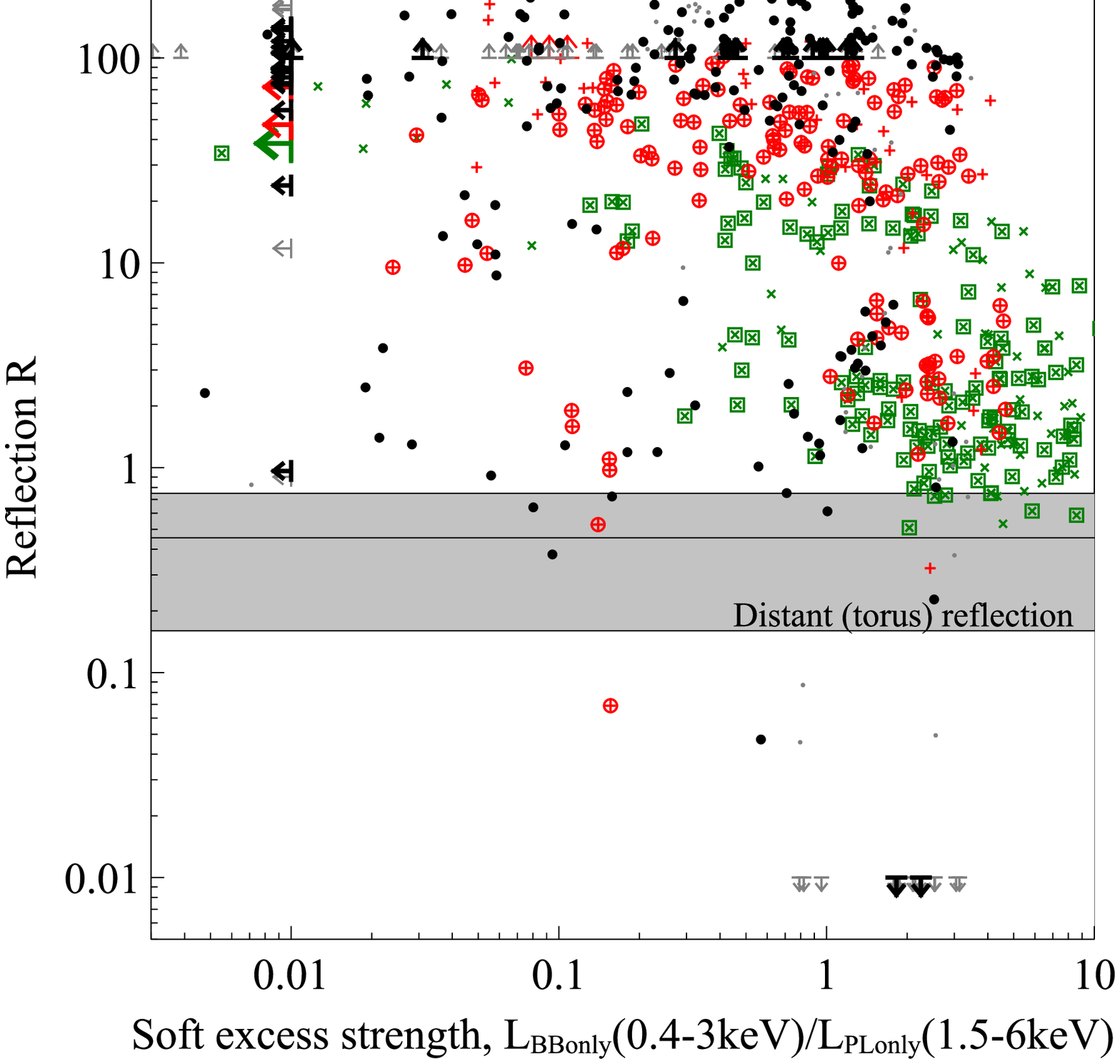}
\caption{Measured hard excess and soft excess strength for the ionised reflection scenario, split by 1) ratio of direct-to-reflected component $N_{\rm pexrav}/N_{\rm reflionx}$ (top panel), and 2) ionisation parameter $\xi$ (lower panel).  In the top panel, small and large black circles represent simulated spectra with $1<N_{\rm pexrav}/N_{\rm reflionx}<10$ (successful fits and 'good' fits respectively); red crosses (successful fits) and red circled crosses (good fits) represent $10<N_{\rm pexrav}/N_{\rm reflionx}<100$ spectra, and green X symbols (successful fits) and green X symbols within squares (good fits) represent $100<N_{\rm pexrav}/N_{\rm reflionx}<1000$ spectra.  In the lower panel, the same sequence of symbol combinations is used to split the reflection-dominated ($1<N_{\rm pexrav}/N_{\rm reflionx}<10$) subset of spectra into groups with $1<\xi<10$, $10<\xi<100$ and $100<\xi<1000$, respectively.  The colours of the upper/lower limit arrows matches the colours of the points for the corresponding parameter groups, and the size of the upper/lower limit symbols increases going from small-to-high values of $N_{\rm pexrav}/N_{\rm reflionx}$ and $\xi$.
\label{fig:ionisedreflsplitbyNratioandxi}}
\end{figure}

This range of $R$ values has not been seen in the real AGN population so far, but only a handful of observations currently exist of reflection-dominated states of AGN (e.g., \citealt{2008MNRAS.391.2003Z, 2012MNRAS.419..116F}). Such prominent hard excesses may be found with new \emph{NuSTAR} observations. Since the ionisation parameter is proportional to the luminosity, the very low luminosity sources (with potential for the most prominent hard excesses) may also be selected out of most X-ray surveys.

For the ionised absorption model, the highest values of the measured $R$ occur for $\rm log(\xi) < 2.75$.  From inspection of the spectra, we find that many of these low ionisation absorbers have absorption troughs that look more like extended `edges'.  The hard portion of the spectrum can then be fit as a pure reflection component in some cases, leading to high values of $R$.  In the remainder of low-ionisation absorber spectra, the slope of the intermediate (3---10~keV) region is too extreme to be fit by the {\sc pexrav} component and the fit fails altogether (discussed further in \S\ref{subsec:failedabsfits}).

We also consider the possibility that the low exposure time (10ks) and resulting poor-quality data at high energies could lead to such anomalous $R$ values.  However, increasing the observation time to 20ks (for both ionised reflection and absorption models, Fig.~\ref{fig:20ks}) did not produce any significant change in the results and the broad trends seen.

\begin{figure}
\includegraphics[width=9.0cm]{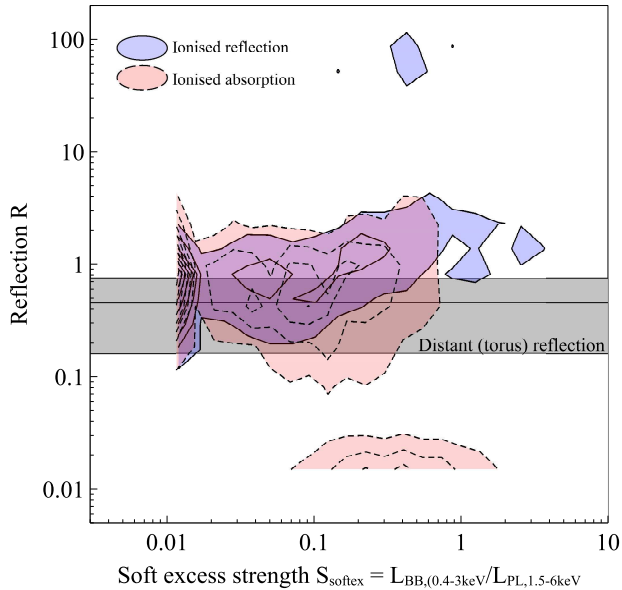}
\caption{Simulation results for the $R-S$ diagram with an observation time of 20~ks.  Key as for Fig.~\ref{fig:refl_vs_softex_ionisedreflsimulations}.
\label{fig:20ks}}
\end{figure}
\subsection{Failed fits with ionised absorption}
\label{subsec:failedabsfits}

As mentioned in \S\ref{subsec:ionabs}, 14\% of the simulated ionised absorption spectra result cannot be fit with the Observer's Model.  We investigate this class of objects in more detail. We find that the only weak discriminant of whether a spectrum fits successfully or not is the ionisation parameter of the absorber, $\rm log(\xi)$.  Failed fits only occur for spectra simulated with $2.1<\rm log(\xi)<2.75$.   There are 1029 possible parameter combinations/simulated spectra in this range (i.e., 3 bins in $\rm log(\xi)$ and 7 bins for each other parameter), out of which 328 model combinations fail completely.  On inspection, these spectra exhibit ionised absorber troughs that look more like edges with a very steep rise in flux towards higher energies.  When manually fit with the observer's model combination, we find that the fits consistently fail.  The power-law regime of the {\sc pexrav} component attempts to fit to the rising part of the edge below 10~keV, but the photon index required is too extreme (see Fig.~\ref{fig:ionisedlowionis_failed}).  This would be a readily identifiable subset of spectral shapes which cannot be fit with a simple black body and reflection model, and any soft `excess' seen would clearly be distinguishable from one produced by ionised reflection.

Of the remaining 701 low-$\rm log \thinspace \xi$ spectra which are successfully fit by the Observer's model, only 50\% qualify as 'good fits' according to the dual-$\chi^{2}$ criteria.   Inspection of these `good fits' reveals that the power-law component is able to fit the rising 1--6~keV portion of the spectrum successfully.  It is therefore not straightforward to identify a very specific part of ionised-absorber parameter space that eludes fitting with the Observer's model, but we can say that the failing of the fit is restricted to low-ionisation absorbers.  The success rate of fitting such unusual spectral shapes may also be subtly dependent on the initial conditions employed in the Observer's model, the exploration of which is beyond the scope of this paper.

\begin{figure}
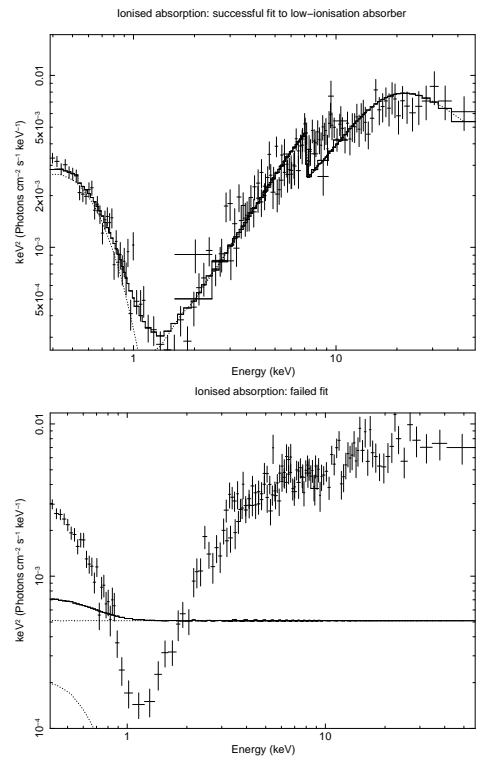

\includegraphics[width=5.0cm,angle=270]{ionisedabs_successfulfit.eps}
\includegraphics[width=5.0cm,angle=270]{ionisedabs_failedfit.eps}
\caption{Example spectra with low-ionisation ionised absorbers, showing the attempted Observer's Model fit using the {\sc pexrav+bbody} model combination. \emph{Top panel}: example of a successful fit.  The input parameters for the simulated spectrum are $\Gamma=2.0$, $N_{\rm H}=5.0 \times 10^{23} \rm cm^{-2}$, $\rm log \thinspace \xi=2.73$, $\sigma = 0.43$.  \emph{Lower panel}: example of a failed fit.  The input parameters are $\Gamma=2.0$, $N_{\rm H}=1.87 \times 10^{23} \rm cm^{-2}$, $\rm log \thinspace \xi=2.42$, $\sigma = 0.23$.
\label{fig:ionisedlowionis_failed}}
\end{figure}

\subsection{Predictions for other comparable instruments}

We have presented results for the \emph{XMM-NuSTAR} instrument combination since \emph{NuSTAR} is newly launched and this instrument combination is already being used to obtain simultaneous broad-band X-ray data on AGN (e.g., \citealt{2013Natur.494..449R,2014arXiv1401.5235M}); therefore it is the most relevant prediction for the current circumstances with real near-term prospects of verifying the work in this paper.  \emph{ASTRO-H} and \emph{ASTROSAT} are also on the horizon, and to check whether our predictions hold for other instruments, we also perform simulations for the ionised reflection model using the current \emph{ASTRO-H} predicted response matrices, assuming the same 10~ks exposure time.   We caution that these responses may be more uncertain than the \emph{XMM+NuSTAR} responses, as real responses are likely to differ significantly from the pre-launch predicted ones.  The resulting $R-S$ diagram does not show any significant difference to that obtained for \emph{XMM-NuSTAR} (Fig.~\ref{fig:astroh}), suggesting that the conclusions in this paper should hold for other similarly-equipped future broad-band X-ray observatories.

\begin{figure}
\includegraphics[width=9.0cm]{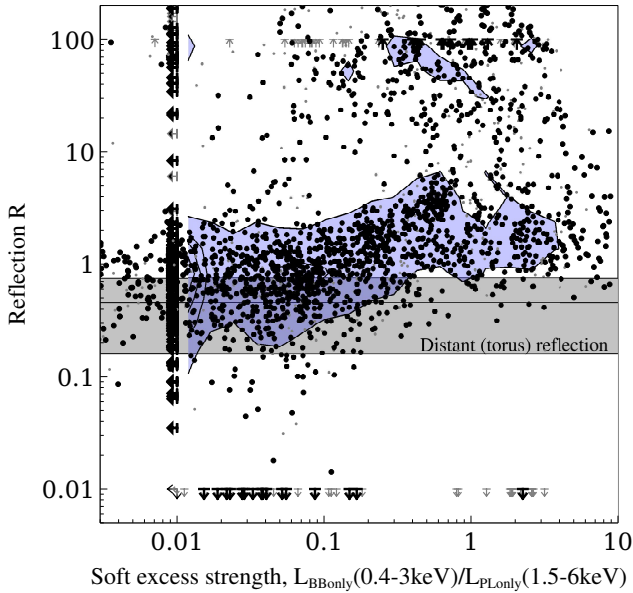}
\caption{Results for ionised reflection using the \emph{ASTRO-H} response matrices.  Key as for Fig.~\ref{fig:refl_vs_softex_ionisedreflsimulations}.
\label{fig:astroh}}
\end{figure}

\subsection{On the feasibility of observing these trends in real AGN samples}

The simulations here outline the trends expected in $R-S_{\rm softex}$ space for two different models using a large grid of 2401 parameter combinations.  AGN samples are typically much smaller due to the competition for observation time, so we require a measure of how feasible it will be to detect these trends in real samples of AGN.  We perform further simulations to estimate the typical AGN sample size required, before a correlation (or the absence of one) can be seen between $R$ and $S_{\rm softex}$.

We use a Monte-Carlo rejection method to simulate $N$ values of $R$ and $S_{\rm softex}$ (corresponding to a sample of $N$ AGN) randomly determined throughout the available $R-S_{\rm softex}$ space, using the contours determined from Figs.~\ref{fig:refl_vs_softex_ionisedreflsimulations} and \ref{fig:refl_vs_softex_ionisedabssimulations} as the probability distribution with which to distribute the points in $R-S_{\rm softex}$ space.  We vary $N$ and measure the resultant Kendall's $\tau$ correlation coefficient of the faked sample to determine when the strength of the correlation approaches that seen in the full simulations.  We do this for both ionised reflection and ionised absorption.  The variation of the correlation coefficient with $N$ for each input model is shown in Fig.~\ref{fig:kendalltau_vs_samplesize}.

\begin{figure}
\centerline{
\includegraphics[width=9.0cm]{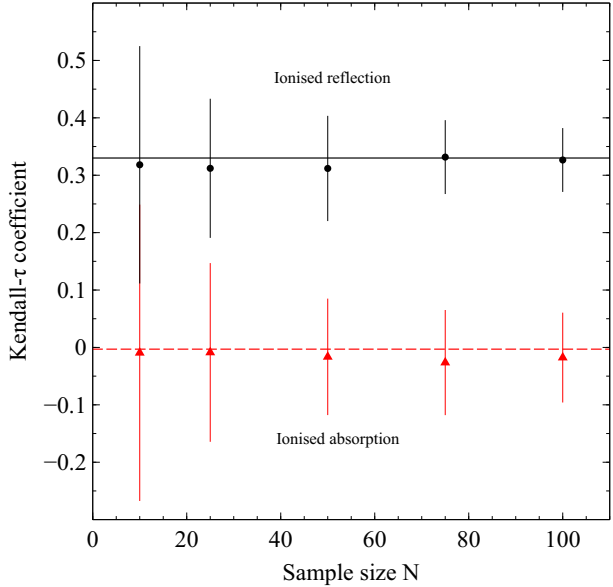}
}
\caption{Kendall's $\tau$ correlation coefficient between $R$ and $S_{\rm softex}$ against sample size, for a simulated AGN sample of size $N$.  The solid (black) and dashed (red) lines show the correlation coefficients measured for the whole set of simulations (corresponding to $N > 1500$), for ionised reflection and ionised absorption respectively.  The error bar shows the standard deviation from 100 Monte-Carlo realisations of such samples of $N$ objects, and indicates that as $N$ approaches $\sim 60$, the correlations exhibited by samples of predominantly reflection-dominated or absorption-dominated samples can be clearly distinguished.
\label{fig:kendalltau_vs_samplesize}}
\end{figure}

We see that at $N \gtrsim 60$, the uncertainty in the correlation coefficients produced by the two processes drops such that the two processes can be distinguished at a 2--3$\sigma$ level.  Therefore, even though the physical process may be ambigious for an individual object based on its location in the $R-S_{\rm softex}$ plane, the trend exhibited by a well-selected, representative AGN sample in this plane can provide an indication of the physics most relevant for the majority of AGN in the sample.

\subsection{Outstanding issues}

We outline some issues to be further explored in future work.  Firstly, the ratio $N_{\rm pexrav}/N_{\rm reflionx}$ is currently used as an estimator of the degree of light bending in the reflection scenario.  A more physical understanding of this parameter is needed to place appropriate upper and lower limits on it for the simulations.  It is easy to conceive of a situation where the direct power-law dominates; however understanding the lower-limit on the range of physically viable ratios (here assumed to be Unity) is more complex and requires a fuller consideration of the energetics of the reflection-dominated state than is undertaken here.  This may impact the strongest observable strengths of the soft excess from the reflection scenario, as initially investigated by \cite{2006A&A...449..493C}, but the observation of AGN in fully reflection-dominated states does support the possibility that $N_{\rm pexrav}/N_{\rm reflionx}$ can take very low values (even $<1$).

We have assumed uniform or log-uniform distributions for the input parameters in these simulations, as the simplest possible scenario.  However, it may be the case that the real underlying distributions are not uniform, or that there are correlations between the input parameters.  There are a handful of studies on this in the literature, presenting the photon index and luminosity distributions in AGN (e.g., \citealt{2011A&A...530A..42C}, \citealt{2013ApJ...763..111V}).  Previous works (e.g.,\citealt{2008ApJ...675.1048R,2011ApJ...734..112B}) have found a range of ionisation parameters for ionised reflection consistent with that assumed here.  There are suggestions of a weak correlation between photon index and luminosity (e.g., \citealt{2008AJ....135.1505S}) hence potential for a correlation between $\Gamma$ and $\xi$.  However, there is no detailed work on the true underlying distributions of physical parameters of reflectors.  It is possible that the observed trends highlighted in this work in the $R-S$ plane could be changed, and the density of points in different parts of the plot could be altered, if different underlying distributions are used.  To allow for this to be investigated in future, we make our simulation results public with the online data accompanying this paper.  The interested researcher can then draw from the provided results in a non-uniform way using more updated distributions or correlations between parameters, as and when such details become more well-constrained.  However, one instinctively expects some degree of correlation between hard- and soft-excess strengths in the ionised reflection scenario, regardless of the precise underlying parameter distributions.

One of the latest models to be proposed for the soft excess is the {\sc optxagnf} model \citep{2012MNRAS.420.1848D}.  Their model combines disk emission, Comptonisation and a power-law in an energetically self-consistent way, where the inner part of the accretion flow below a \emph{coronal radius} is Comptonised to produce the soft excess. Comptonisation is the engine behind the soft excess in this model, and one does not expect it to exert any influence on the hard X-ray emission based on simple models (e.g., \citealt{2004MNRAS.352..523P}) providing a purely standalone component for the soft excess.  Therefore, one does not expect any link between the soft excess and hard excess in this scenario.  However, this may not be the case for {\sc optxagnf} where the disc and corona geometries are linked by energetic considerations, and we are currently undertaking a study of this model in detail, with a view to produce a similar $R-S_{\rm softex}$ diagnostic plot to compare it with the models in this paper.  The {\sc optxagnf} model is able to fully reproduce the optical--to--X-ray SED up to 10~keV, as shown by the comprehensive study of \cite{2012MNRAS.420.1825J}, but its hard X-ray signatures have not been studied.  Additionally, distant reflection needs to be added in a consistent fashion if we want to compare it with the models studied here, where {\sc pexrav} was used to provide the direct continuum along with the distant reflection.  How to do this is not clear, and since {\sc optxagnf} has many more model parameters to consider than the models in this paper, we defer this study to a future paper (in prep.).

\section{Summary}
\label{sec:summary}

This work points to a scheme whereby different candidate models for the soft excess can be distinguished in a plot of measured reflection against soft-excess strength, assuming the simplest possible {\sc pexrav + bbody} model combination is fit to the spectra, according to the scheme in Fig.~\ref{fig:refl_vs_softex_bothmodels_summary}.  This methodology can readily be extended to other candidate models, and we are currently in the process of producing such a diagnostic for the {\sc optxagnf} model, where photons from the inner part of the accretion disc are Comptonised to produce the soft excess \citep{2012MNRAS.420.1848D}.  A key advantage of this approach is its economy: data of moderate quality can be used, gathered using short exposures (e.g. 10~ks in both \emph{XMM} and \emph{NuSTAR} for a source of the brightness of NGC 4051) to gain real physical insight into the energy production mechanisms in AGN, \emph{without} requiring fitting of more complex models to long-exposure, very high signal-to-noise ratio data.  This approach will therefore be useful in constraining the soft excess mechanism in samples of AGN, where it may be challenging to obtain such long exposures on each source.  This approach is particularly useful for samples of AGN where trends can be discerned, although it can be used for individual AGN as well, if they lie in unambigious regions of $R-S_{\rm softex}$ space.  Notably, ionised reflection predicts a clear relation between $R$ and $S_{\rm softex}$, but ionised absorption does not.

As simultaneous or broad-band X-ray data comes in from \emph{NuSTAR}+\emph{XMM}, \emph{ASTRO-H} and \emph{ASTROSAT}, this plot can be populated with accurate, simultaneous determinations of the strengths of the hard and soft excesses in samples of real AGN, updating the work presented in Fig.~\ref{reflvssoftex_BAT58monAGN}.  Using the contours presented in Fig.~\ref{fig:refl_vs_softex_bothmodels_summary} as probability contours, we simulate a smaller sample of AGN using a Monte-Carlo Rejection method, and estimate that $\sim 60$ AGN would be sufficient to verify the existence of a $R-S_{\rm softex}$ correlation of comparable strengths to those found in our original simulations (Figs~\ref{fig:refl_vs_softex_ionisedreflsimulations} and \ref{fig:refl_vs_softex_ionisedabssimulations}).  This would amount to an easily achievable \emph{XMM}-\emph{NuSTAR} campaign of $\sim$600~ks.

It will be easier to narrow down the mechanism responsible for producing the soft excess in any given source, if it is first fit with the simplest possible ``observer's model'' outlined here to locate it on the $R-S_{\rm softex}$ plot.  Additionally, the regions of this plot occupied by large samples of AGN will also provide an indication of the most likely soft excess production mechanisms in the AGN population as a whole.  The general simulation methodology adopted here also has much potential for distinguishing between competing models in other areas of both AGN science and other fields of astrophysics.

\section{Acknowledgements}
\label{sec:ack}
We thank the anonymous referee for useful suggestions which improved the paper.  CSR thanks NASA for support under grant NNX12AE13G. We thank Jeremy Sanders for help with the use of his \emph{Veusz} plotting package, and Cole Miller for helpful discussions on Monte-Carlo simulation techniques.

\bibliographystyle{apj} 
\bibliography{softexcessreflection}

\end{document}